\documentclass[manuscript]{aastex}

\usepackage{graphicx,pdflscape,color,url}

\newcommand{\Msun}{\ensuremath{\mathrm{M}_\odot}}
\newcommand{\Abu}[1]{\ensuremath{A(\mathrm{#1})}}
\newcommand{\dex}{\ensuremath{\mathrm{dex}}}
\newcommand{\B}{\ensuremath{\mathrm{B}}}
\newcommand{\A}{\ensuremath{\mbox{\AA}}}
\newcommand{\R}[2]{\ensuremath{\mathrm{#1}/\mathrm{#2}}}
\newcommand{\SB}[1]{\ensuremath{[#1]}}
\newcommand{\SR}[2]{\SB{\R{#1}{#2}}}
\newcommand{\SH}[1]{\SR{#1}{H}}

\shorttitle{Oxygen abundance of  SM0313$-$6708}
\shortauthors{Bessell, Collett, Keller, et al.}

\begin{document}

\title{Nucleosynthesis in a Primordial Supernova:
Carbon and Oxygen Abundances in SMSS J031300.36$-$670839.3\thanks{Based on observations obtained with European Southern Observatory (ESO) telescopes (proposal 092.D-0742)}}
%

\author{Michael S. Bessell}
\affil{Research School of Astronomy and Astrophysics, CPMS, The Australian National University, Canberra, ACT 2611, Australia}
\email{michael.bessell@anu.edu.au}

\author{Remo Collet}
\affil{Research School of Astronomy and Astrophysics, CPMS, The Australian National University, Canberra, ACT 2611, Australia}
\email{remo.collet@anu.edu.au}

\author{Stefan C. Keller\altaffilmark{1}}
\affil{Research School of Astronomy and Astrophysics, CPMS, The Australian National University, Canberra, ACT 2611, Australia}
\email{stefan.keller@defence.gov.au}

\author{Anna Frebel}
\affil{Department of Physics and Kavli Institute for Astrophysics and Space Science, Massachusetts Institute of Technology, Cambridge, MA 02139, USA}
\email{afrebel@mit.edu}

\author{Alexander Heger\altaffilmark{2,3}}
\affil{Monash University, School of Physics and Astronomy, Victoria 3800, Australia}
\email{alexander.heger@monash.edu}

\author{Andrew R. Casey}
\affil{Institute of Astronomy, University of Cambridge, Cambridge CB3 0HA, United Kingdom}
\email{arc@ast.cam.ac.uk}

\author{Thomas Masseron}
\affil{Institute of Astronomy, University of Cambridge, Cambridge CB3 0HA, United Kingdom}
\email{tpm40@ast.cam.ac.uk}

\author{Martin Asplund}
\affil{Research School of Astronomy and Astrophysics, CPMS, The Australian National University, Canberra, ACT 2611, Australia}
\email{martin.asplund@anu.edu.au}

\author{Heather R. Jacobson}
\affil{Kavli Institute for Astrophysics and Space Science, Massachusetts Institute of Technology, Cambridge, MA 02139, USA}
\email{hrj@mit.edu}

\author{Karin Lind}
\affil{Department of Physics and Astronomy, Uppsala University, Box 516, SE-751 20, Uppsala, Sweden}
\email{karin.lind@physics.uu.se}

\author{Anna F. Marino}
\affil{Research School of Astronomy and Astrophysics, CPMS, The Australian National University, Canberra, ACT 2611, Australia}
\email{anna.marino@anu.edu.au}

\author{John E. Norris}
\affil{Research School of Astronomy and Astrophysics, CPMS, The Australian National University, Canberra, ACT 2611, Australia}
\email{john.norris@anu.edu.au}

\author{David Yong}
\affil{Research School of Astronomy and Astrophysics, CPMS, The Australian National University, Canberra, ACT 2611, Australia}
\email{david.yong@anu.edu.au}

\author{Gary Da Costa}
\affil{Research School of Astronomy and Astrophysics, CPMS, The Australian National University, Canberra, ACT 2611, Australia}
\email{gary.dacosta@anu.edu.au}

\author{Conrad Chan}
\affil{Monash University, School of Physics and Astronomy, Victoria 3800, Australia}
\email{conrad.chan@monash.edu}

\author{Zazralt Magic}
\affil{Max-Planck-Institut f\"ur Astrophysik, 85741 Garching, Germany}
\email{magic@MPA-Garching.MPG.DE}

\author{Brian Schmidt}
\affil{Research School of Astronomy and Astrophysics, CPMS, The Australian National University, Canberra, ACT 2611, Australia}
\email{brian.schmidt@anu.edu.au}

\author{Patrick Tisserand\altaffilmark{4}}
\affil{Sorbonne Universit\'es, UPMC Univ Paris 06, UMR 7095, Institut d'Astrophysique de Paris, F-75014, Paris, France}
\email{tisserand@iap.fr}


\altaffiltext{1}{Present address: Department of Defense, Australian Government, Canberra, Australia}
\altaffiltext{2}{University of Minnesota, School of Physics and Astronomy, Minneapolis, MN 55455, USA}
\altaffiltext{3}{Shanghai Jiao-Tong University, CNA, Department of Physics and Astronomy, Shanghai 200240, P.~R.~China}
\altaffiltext{4}{UPMC-CNRS, UMR7095, Institut d'Astrophysique de Paris, F-75014, Paris, France}


\begin{abstract}
SMSS J031300.36$-$670839.3 (hereafter SM0313$-$6708) is a sub-giant
halo star, with no detectable Fe lines and large overabundances of C
and Mg relative to Ca.  We obtained VLT-UVES spectra extending to
$3060\,\A$ showing strong OH A-X band lines enabling an oxygen
abundance to be derived.  The OH A-X band lines in SM0313$-$6708 are
much stronger than the CH C-X band lines. Spectrum synthesis fits
indicate an [$\R{O}C$] ratio of $0.02\pm 0.175$.
Our high S/N UVES data also enabled us to lower the Fe abundance
limit to $\SH{Fe}_\mathrm{\langle3D\rangle,NLTE}<-7.52$ ($3\,\sigma$).
These data support our previous suggestion that the
star formed from the iron-poor ejecta of a single massive star
Population III supernova.
\end{abstract}

\section{Introduction}
\citet{Kell14} described the discovery of SMSS J031300.36$-$670839.3
(hereafter SM0313$-$6708) and the analysis of the high
resolution spectrum taken with the Magellan Inamori Kyocera Echelle
\citep[MIKE:][] {Bern03} spectrograph on the $6.5\,$m Magellan
Telescope at Las Campanas Observatory in Chile.  The spectrum was
remarkable for the complete absence of lines of iron or any other
metal, apart from Mg$\,$I and Ca$\,$II.  The
fact that no iron line is evident in the signal-to-noise (S/N)
$\sim100$ (at $\lambda = 4000\,\A$) spectrum indicated that the iron
abundance in SM0313$-$6708 was at most $\SH{Fe}=-7.2$,%
\footnote{logarithm of abundance ratio relative to present day solar
  surface abundances,
  $\mathrm{[A/B]}=\log(n_\mathrm{A}/n_\mathrm{B})-\log(n_\mathrm{A}/n_\mathrm{B})_\odot$}
making this star at least ${\sim}40$ times more iron-poor than HE
1327-2326 \citep[$\SH{Fe}=-5.6$;][]{Freb05}.

The only elements whose abundances could be measured were lithium,
carbon, magnesium, and calcium with
$\Abu{Li}=12+\log(\R{Li}H)=0.71\pm0.10$, $\SH{C}=-2.44\pm0.10$,
$\SH{Mg}=-4.30\pm0.10$, and $\SH{Ca}=-7.2\pm0.10$, respectively.  The
level of lithium is presumably related to the decline from the
primordial Spite plateau level of $\Abu{Li}=2.2$ during early post
main sequence stellar evolution (Lind et al.~2009) and this evidence,
together with a $T_\mathrm{eff}=5125\,$K and $\log g=2.3$ derived from
spectrophotometry, is consistent with a star ascending the giant
branch for the first time.  The C, Mg, and Ca abundances can be
attributed to a low energy Pop III SNe and measurement of the oxygen
abundance, in particular, offers an important test of this scenario.
To measure the oxygen abundance in ultra-metal poor K stars, however,
it is necessary to work in the ultra-violet and observe the OH lines
that persist well after the [OI] $6300.3\,A$ line has become
undetectable.

\section{Observations}

In order to measure the OH (and CH) lines between $3100\,\A$ and
$3200\,\A$, and obtain better detection limits on N, Na, V, Co, Ni,
Fe, and Cu, we used UVES (Ultraviolet and Visual Echelle Spectrograph;
\citealt{Dekk00}) on the VLT (ESO Very Large Telescope) at Paranal in
Chile. Between September 2013 and September 2014, 31 exposures
totalling $29.26$ hours of on-source integration were obtained. The
UVES UV spectrum extended from $3060\,\A$ to $3860\,\A$ and the two
red spectra from $4786\,\A$ to $6803\,\A$.

\begin{figure*}
  \centering
 \includegraphics{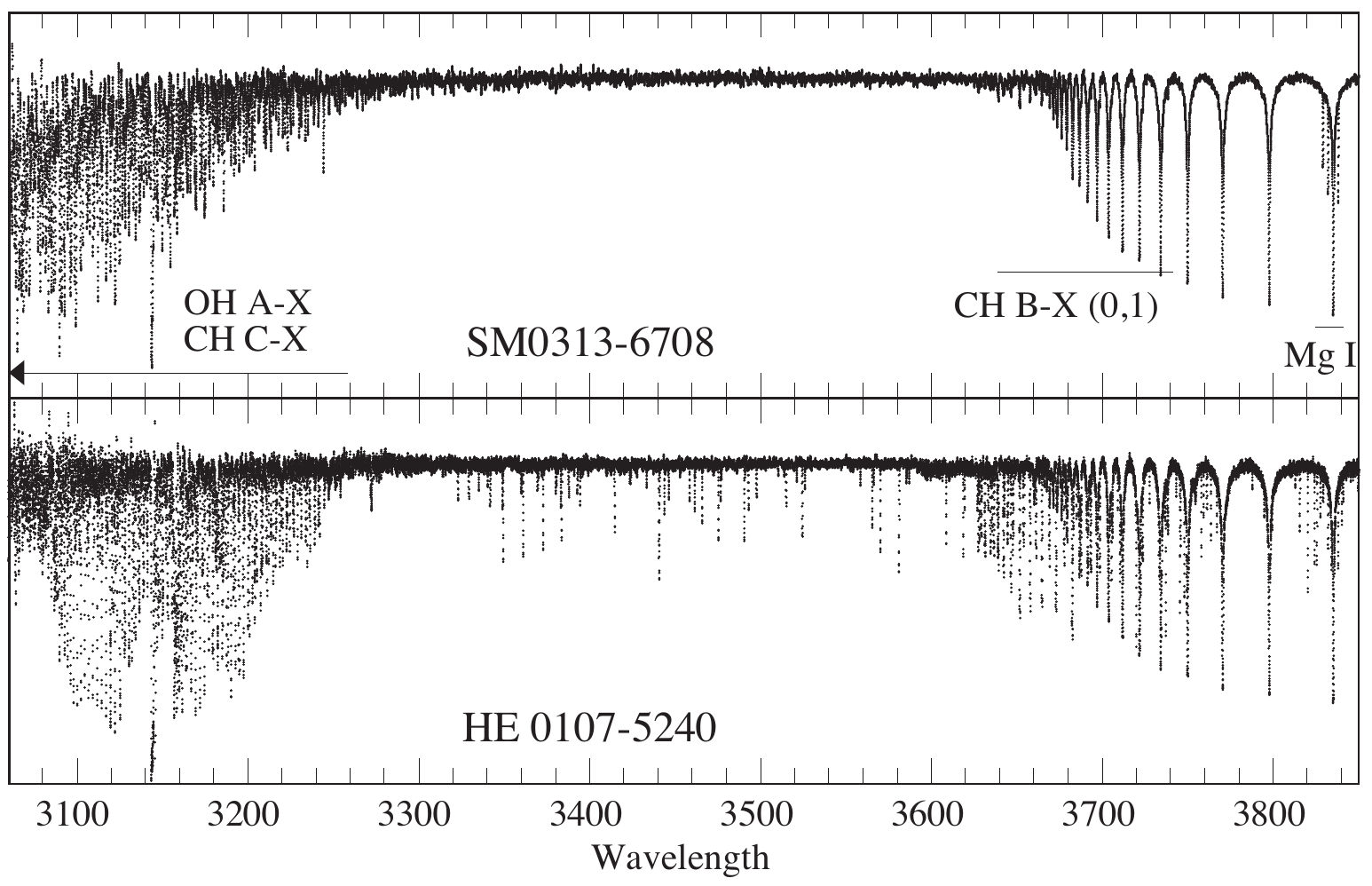}
   \caption{The normalized UVES UV spectra of
     SM0313$-$6708 and HE 0107$-$5240. Note the different
     distribution of the OH and CH lines in the two stars and the
     complete absence of metal lines in SM0313$-$6708 except for the
     similar strength Mg\,I lines.}
  \label{fig:UVES}
\end{figure*}

The data were reduced using the default settings of the UVES pipeline
as implemented in the ``reflex'' reduction environment. There was a
high order sinusoidal-like modulation of the continuum remaining in
much of the UV spectrum. The modulation with an amplitude of about
$20\,\%$ and frequency about $50\,\A$ was removed by hand, leaving a
residual variation of amplitude of $2\,\%-4\,\%$ and spacing of about
$1.5\,\A$ that makes the measurement of shallow features somewhat
uncertain.  It was very clear, however, that there were many strong OH
lines present between the observational cutoff around $3060\,\A$ and
$3200\,\A$, and that the OH lines were much stronger than the CH
lines, except at the band head of the CH C-X band near $3144\,\A$.  In
Fig.~\ref{fig:UVES} we compare the UV spectra of SM0313$-$6708
($\SH{Fe}\le -7.2$) and HE 0107$-$5240 ($\SH{Fe}=-5.4$), a C-rich ultra-EMP
star with almost identical effective temperature and gravity, and possibly at
the same evolutionary stage.
The total absence of metal lines in SM0313$-$6708 (except for the Mg\,I
lines around $3830\,\A$) is very marked as is the different
distribution of the forest of molecular lines below $3250\,\A$.  The
CH lines dominate the spectrum of HE 0107$-$5240 whereas OH lines
dominate in SM0313$-$6708.  The complete absence of absorption lines in
SM0313$-$6708 between $3250\,\A$ and $3800\,\A$ enables more stringent
limits to be placed on those species with strong lines in that region.

\section{Analysis}
We have carried out a spectrum synthesis of the CH, OH, and NH
molecular lines in SM0313$-$6708 using Turbospectrum
\citep{Alva98,Plez12}. We adopted an interpolated ``standard'' metallicity MARCS 1D LTE
model atmosphere structure \citep{Gust08}, for $T_\mathrm{eff}=5125\,$K,
$\log g = 2.30$, $\SH{M}=-5.0$,  but with actual abundances
from this paper for the atomic and molecular equilibria
calculations. We used the new $^{12}$CH and $^{13}$CH line list from
\citet{Mass14}, while for OH and NH we used extensive line lists
recently recomputed by Masseron (in prep.).  The Masseron OH $g\!f$
values are very similar to those of \citet{Gill01} but the NH $g\!f$
values are significantly smaller by $0.3 - 0.7\,\dex$ than the
commonly used values of \citet{Kur11}. The solar
composition was adopted from \citet{Aspl09}.

We refitted the CH A-X and B-X lines in the blue MIKE spectrum and
determined a carbon abundance of $\mathrm{[C/H]}=-2.41\pm0.05$, very similar
to the previously determined A-X value. The CH C-X band synthesis
is in good agreement with the A-X and B-X bands.  The synthesised
spectrum well reproduces the observed CH lines in strength and
position for all $J$ values and for all branches observed.

\begin{figure}
  \centering
  \includegraphics{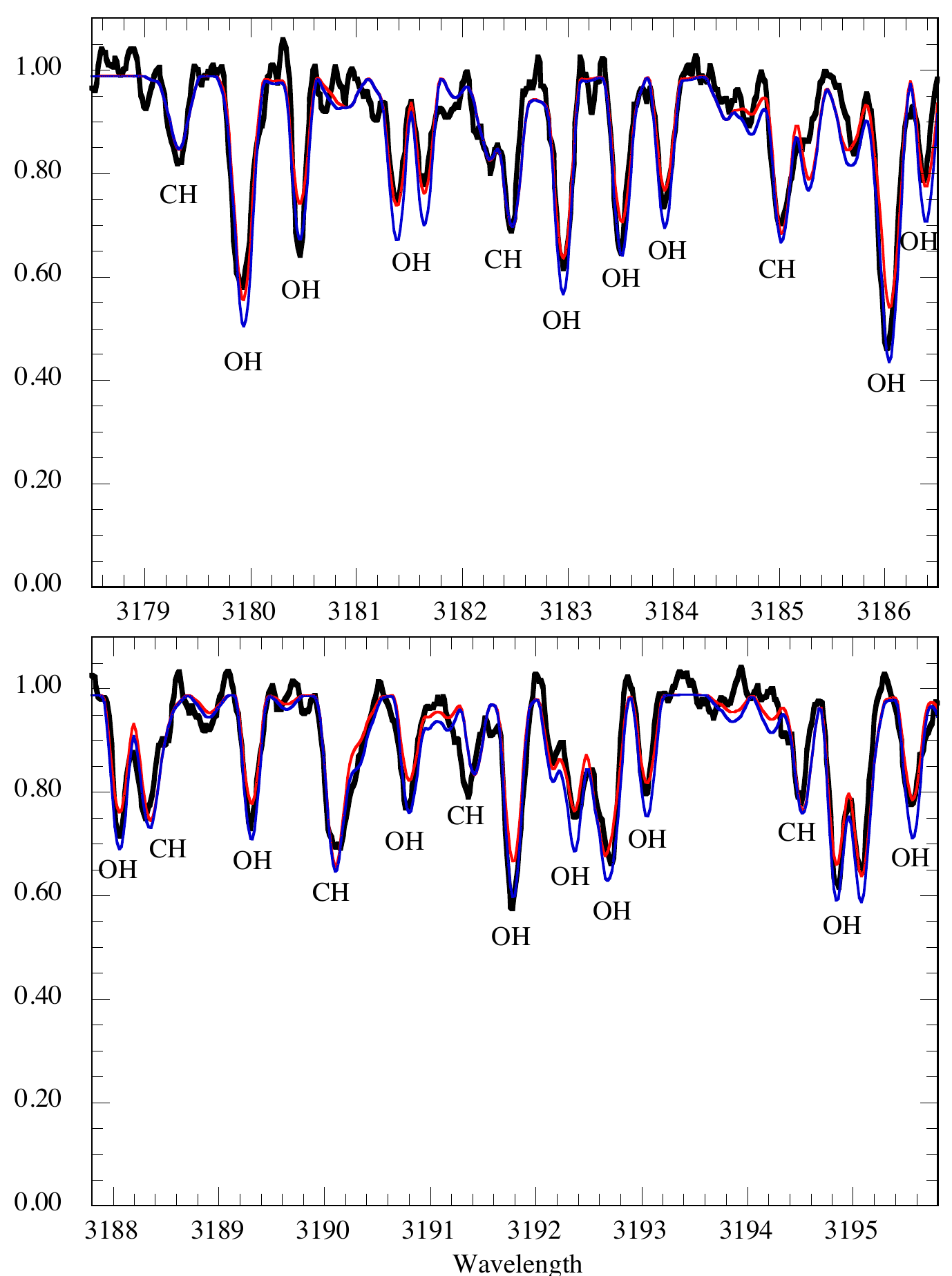}
      \caption{Region between $3178\,\A$ and $3196\,\A$ in
        SM0313$-$6708.  Observations (thick black line); synthetic
        spectra for $\SH{C}=-2.41$ and oxygen abundance
        $\SH{O}=-2.10$ (red line) and higher abundance $\SH{O}=-1.90$
        (blue line).  Although the OH lines here are weaker than the
        $3060\,\A-3200\,\A$ region, the S/N is higher and the
        continuum level is better defined.}
  \label{fig:3178}
\end{figure}

\begin{figure*}
  \centering
  \includegraphics[width=\textwidth]{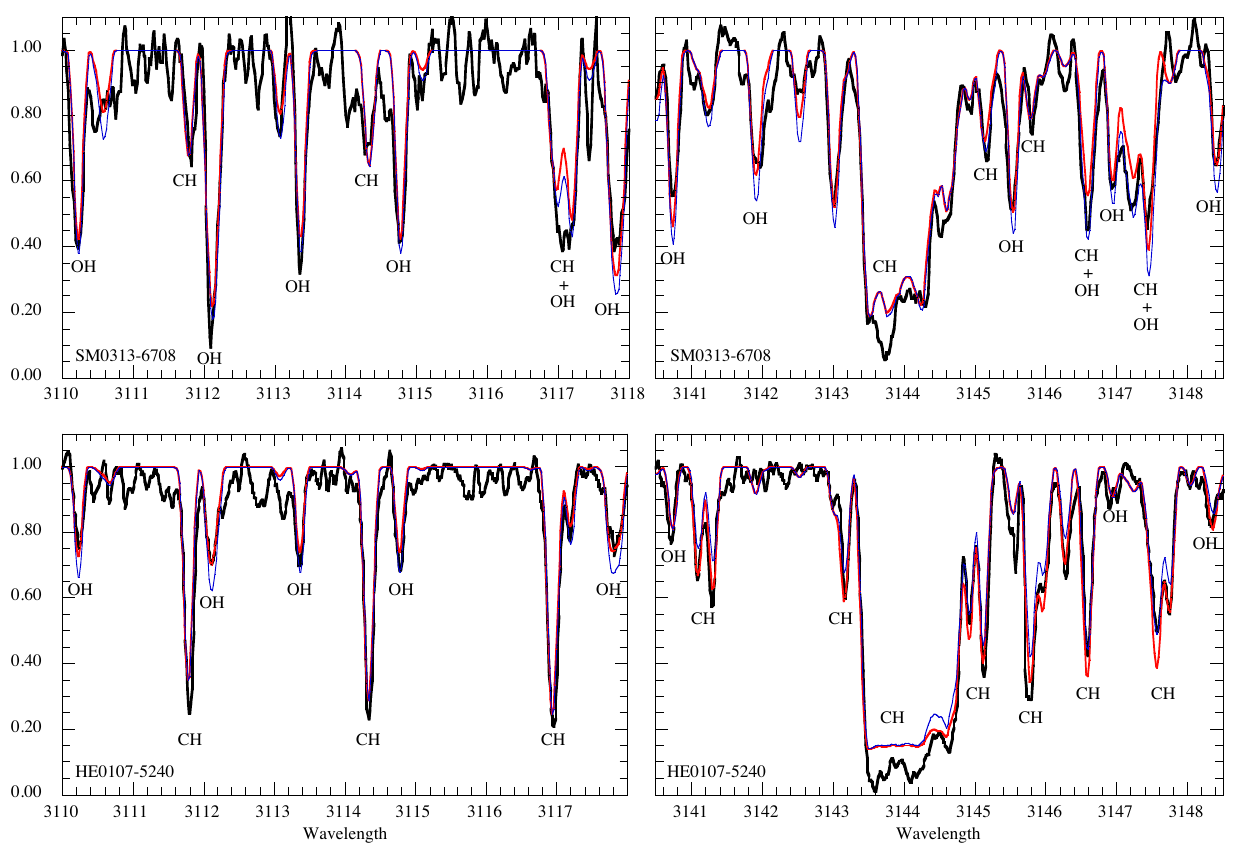}
    \caption{A spectral region showing mainly OH lines (\textbf{left})
      and a region around the CH C-X band head
      (\textbf{right}). \textbf{Top:} SM0313$-$6708. Observations (thick
      black line); synthetic spectra for abundance $\SH{O}=-2.10$ (red
      line) and higher abundance $\SH{O}=-1.90$ (blue line).
      $\SH{C}=-2.41$ in both syntheses.  The $3110\,\A-3118\,\A$
      region has low S/N and uncertain continuum placement.
      \textbf{Bottom:} HE 0107$-$5240.  Observations (thick black line);
      \textbf{left:} synthetic spectra for best fitting abundances
      $\SH{C}=-1.20$, $\SH{O}=-2.55$ (red line) and higher abundance
      $\SH{O}=-2.40$ (blue line).  \textbf{Right:} $\SH{O}=-2.55$,
      $\SH{C}=-1.20$ (red line) and lower abundance $\SH{C}=-1.40$
      (blue line).}
  \label{fig:3110}
\end{figure*}

The OH lines extend from $3060\,\A$ to redward of $3200\,\A$.  Because
of the lower S/N and the uncertain continuum placement at the shortest
wavelengths where the lines are strongest, we derived the oxygen
abundance in SM0313$-$6708 by fitting the weaker OH lines between
$3178\,\A$ and $3200\,\A$.  Fig.~\ref{fig:3178} shows clearly that the
oxygen abundance is well constrained at $\SH{O} =-2.00\pm0.10$.  In HE
0107-5240, the similar temperature and gravity star with abundances
$\SH{M}\approx-5.4$, $\SH{C}=-1.58$ \citep{Chris04} and $\SH{O}=-3.0$ \citep{Bess04},
these longer wavelength regions could not be used for OH fitting as they are
dominated by CH lines as seen in Fig.~\ref{fig:3110}.  We note that the treatment of
scattering (as done in Turbospectrum), is very important for the analysis of the UV lines
in SM0313-6708 and metal-poor giants in general.

Fig.~\ref{fig:3110} shows the comparison between the observed and
synthetic spectra at the shorter wavelengths in SM0313$-$6708 and
HE 0107$-$5240.  The panels on the left show the
region that was used to measure the oxygen abundance in HE
0107-5240, and the panels on the right shows the region around the CH
C-X band head.
Note the significantly different relative strengths of OH and CH lines
in the two stars.

There is no indication of any NH features centered on $3360\,\A$, but the higher S/N of
the UVES spectrum enables us to reduce the upper limit to
$\SH{N}<-4.2$, at least two orders of magnitude lower than C and O.

\subsection{3D$-$1D corrections to C and O abundances}
The carbon, nitrogen and oxygen abundances reported above were derived
from 1D LTE analysis of the CH and OH lines.  One of us (RC) generated
a 3D time-dependent hydrodynamic model atmosphere of a red giant star
with stellar parameters very similar to the ones of SM0313$-$6708
using a custom version of the {\sc Stagger} code
\citep{nordlund95,nordlund09} as well as a 1D model atmosphere
corresponding to the same parameters and relying on the same
micro-physics and opacity data.  Synthetic LTE CH and OH spectra for a
range of C and O abundances were produced using the {\sc SCATE} code \citep{Hayek:2011}
with 3D and 1D models, accounting for the effects of radiative scattering;
the best fits to the observed spectrum were determined in both cases
and the 3D$-$1D abundance corrections were then derived.  These
3D$-$1D corrections are: $-0.14\,\dex$ for the carbon abundance as
derived from the CH band and $-0.53\,\dex$ for oxygen from OH (the
negative sign meaning the 3D-derived abundances are lower).  The
3D-corrected [$\R{O}C$] ratio in SM0313$-$6708 is
therefore $+0.02$ with an uncertainty of $0.175\,\dex$.  A 3D$-$1D
correction of $-0.50\,\dex$ to the 1D nitrogen abundance was also
derived with similar 3D and 1D LTE calculations but based purely on
the comparison of synthetic 3D and 1D spectra.

The poorer fit of the deep OH and CH lines in Fig.~\ref{fig:3110} is
likely due to limitations with the adopted 1D model
atmosphere. Our computations of CH, NH, and OH lines have
been carried out in LTE, but in low-density atmospheres it is possible
that non-LTE processes may be important for cooling the outer stellar
layers \citep{Lamb13} and for affecting the strength of molecular
bands.  As no study of non-LTE CH, NH, and OH line formation is
available in the literature, however, we are unable to quantify
non-LTE effects on CNO abundances derived from these molecular bands.

\subsection{$^{12}$C/$^{13}$C ratio}
In order to put limits on the $\R{^{12}C}{^{13}C}$ ratio we used two
regions of the spectrum where the $^{12}$CH and $^{13}$CH lines are
well separated and free from blending.  There were five strong lines
in the B-X region and nine in the A-X region.  Sections of spectrum
centered on each of the lines were extracted and summed in both the
observed spectrum and the synthetic spectra.  The summed observed and
synthetic $^{12}$CH lines were in good agreement but no $^{13}$CH
feature was evident in the summed spectrum. This allows us to place a
three sigma lower limit of about $40$ for the $\R{^{12}C}{^{13}C}$
ratio.  This low value of $^{13}$C is consistent with massive
Population III star yields with no or little primary nitrogen
production \citep{HW10}.

\subsection{Revised upper limits on the abundances of Fe, Na, Si, Sc, V, Co, Ni, Cu}
Two of the Fe\,I transitions with the largest $\log g\!f$ values in the
observed spectral range are located within the highest S/N portion of
the UV spectrum, at $3719.94\,\A$ and $3737.13\,\A$.
Sections around the position of
these lines were extracted, summed and compared with the summed
synthetic spectra computed for a range of Fe abundances.  Again, no
feature was evident in the observed spectrum and we estimate a
revised abundance limit for $\SH{Fe}$ of about $-7.5$.
%

The non-detection of lines of Sc, Ti, V, Co, Ni, and Cu in the UVES spectrum
in the range from $3100$~{\AA} to $3700$~{\AA} provided revised
lower upper limits for many elements given in Table~\ref{tab:abu} (Frebel et al., in prep.)
%
%
The non-detection of the
Si\,I line at $3905.52\,\A$ in the blue MIKE spectrum indicates an
upper limit of $-5.6\,\dex$.

The UVES red spectrum was in good agreement with the MIKE spectrum for
the Mg\,I lines and the Li\,I line, but the better S/N and weaker
telluric lines in the UVES red spectrum indicates that the Na limit
could be lowered to about $-5.7\,\dex$.  Table~\ref{tab:abu} lists the
current best estimates for the chemical abundances of SM0313$-$6708.

\section{Discussion}

\citet{HW10} have followed the evolution and explosion of metal-free
stars with masses $10\,\Msun$ to $100\,\Msun$ and determined their
nucleosynthetic yields, light curves, and remnant masses. Such stars
would have been the first to form after the Big Bang and could have
left a distinctive imprint on the composition of the early
universe. The evolution of these stars may be affected in several ways
by their zero-metallicity primordial composition. Because these stars
suffered little mass-loss during their pre-explosive
lifetimes, their masses at their deaths were larger than more
metal-rich stars.  The models produces some carbon, but for masses
less than about $40\,\Msun$ primary production of nitrogen is
suppressed.  These stars ended their lives as compact blue supergiants
making their envelope more tightly bound, with two interesting
consequences: \textit{i)} The more compact envelope increases the
amount of mass that falls back after the explosion, meaning that they
were more likely to collapse to black holes at lower masses compared
to modern stars, trapping most of the heavy elements; \textit{ii)}
mixing by the Rayleigh-Taylor instabilities is suppressed in these
compact stars, so it is mainly unmixed material that escapes in the
explosion.  \citet{HW10} fitted their predicted nucleosynthetic yields
to several examples of metal-poor stars and found that the best fits
were obtained from relatively low energy explosions of $0.6-1.8\,\B$
($1\,\B= 10^{51}\,$erg) as compared to hypernova models
\citep{Naka01}.  Overall, the large sample of metal-poor stars in
\citet{Cayr04} was well fit with a normal Salpeter IMF including all
stars from $10\,\Msun$ to $100\,\Msun$ (but with a preference for
stars lighter than $15\,\Msun$), a small amount of fallback and with
reduced mixing of $1\,\%$ of the helium core mass (for reference, the
default mixing value adopted to solar metallicity stars is $10\,\%$ of
the helium core mass; \citealt{HW10}).  For the two ultra-iron-poor,
but high $\R{O}{Fe}$, stars HE 0107$-$5240 \citep{Chris04,Bess04} and
HE 1327$-$2326 \citep{Aoki06,Freb06}, the best fits were for similar
low energy explosions, a lower upper mass range ($10\,\Msun$ to less
than $30\,\Msun$), small fallback and an even lower mixing of $0.025$.

The abundances of SM0313$-$6708 are much more extreme than the
\citet{Cayr04} EMP stars or the two C-rich ultra-EMP stars.  The biggest
difference are the large carbon, oxygen and magnesium overabundances
relative to calcium, and the non-detection of all other metals.  Of
particular significance is the oxygen to carbon ratio in SM0313$-$6708
of  $\SR{O}C_\mathrm{3D} = +0.02$ compared to
$\SR{O}C_\mathrm{3D} \approx-1.0$ \citep{Collet06}
in HE 0107$-$5240.
\begin{table}

\caption{Chemical abundances of SM0313$-$6708.\label{tbl-1}}
\vspace{2mm}
\begin{tabular}{clllll}
\tableline\tableline
Element &
$\SH{X}_\mathrm{1D,LTE}$ &
$\SH{X}_\mathrm{1D,NLTE}$ &
$\SH{X}_\mathrm{\langle3D\rangle,LTE}$ &
$\SH{X}_\mathrm{\langle3D\rangle,NLTE}$ &
$\SH{X}_\mathrm{3D,LTE}$
\\
\tableline
\
Li\,I \tablenotemark{\ast}      &$0.70\pm0.10$ &$0.73 \pm0.10$ & 	& $0.56\pm0.10$&$\mathbf{0.71\pm0.10}$\\
\tableline
C (CH) \tablenotemark{\ast\ast}	& $-2.41\pm0.05$  &&&& $\mathbf{-2.55\pm0.09}$	\\
N (NH) \tablenotemark{\ast\ast}	& $<\!\!-4.2 $    &&&& $\mathbf{<\!\!-4.70 }$	\\
O (OH) \tablenotemark{\ast\ast}	& $-2.00\pm0.10 $ &&&& $\mathbf{-2.53\pm0.15 }$	\\
Na\,I   \tablenotemark{\ast\ast}	& $<\!\!-5.7 $  & $<\!\!-5.65 $	& $<\!\!-5.79 $& $\mathbf{<\!\!-5.66 }$	\\
Mg\,I  \tablenotemark{\ast\ast}	& $-4.45\pm0.07 $	& $-4.03\pm0.03 $  & $-4.53\pm0.09$ 	& $\mathbf{-4.08\pm0.03}$\\
Al\,I    \tablenotemark{\ast\ast} & $\mathbf{<\!\!-6.2 }$ \\
Si\,I    \tablenotemark{\ast\ast} & $\mathbf{<\!\!-5.6 }$ \\
Ca\,II 	&$-7.46\pm0.10 $	&$-7.27\pm0.10  $ &	$-7.60\pm0.10 $	& $\mathbf{-7.26\pm0.10}$\\
Sc\,II\tablenotemark{\ast\ast}&$\mathbf{<\!\!-5.5 }$\\
Ti\,II\tablenotemark{\ast\ast}&$\mathbf{<\!\!-7.1 }$\\
V\,II\tablenotemark{\ast\ast}&$\mathbf{<\!\!-4.3 }$\\
Cr\,I &$\mathbf{<\!\!-6.3 }$\\
Mn\,I &$\mathbf{<\!\!-5.8 }$\\
Fe\,I \tablenotemark{\ast\ast}&$<\!\!-7.80 $ &$<\!\!-7.28 $ & $<\!\!-8.27 $& $\mathbf{<\!\!-7.52 }$\\
Co\,I \tablenotemark{\ast\ast}&$\mathbf{<\!\!-5.4 }$\\
Ni\,I &$\mathbf{<\!\!-6.9 }$\\
Cu\,I \tablenotemark{\ast\ast}&$\mathbf{<\!\!-5.5 }$\\
Zn\,I &$\mathbf{<\!\!-3.4 }$\\
Sr\,II&$\mathbf{<\!\!-6.9 }$\\
Ba\,II&$\mathbf{<\!\!-6.1 }$\\
Eu\,II&$\mathbf{<\!\!-2.9 }$\\
\end{tabular}
\tablenotetext{\ast}{The abundance of Li is expressed as
  $A(\mathrm{Li})={\log}(N(\mathrm{Li})/N(H)) + 12$}
\tablenotetext{\ast\ast}{Abundances derived in this paper. }
\tablenotetext{}{ Estimated ($1\sigma$) observational uncertainties in
  the quoted upper limit abundances are $0.3\,\dex$. For Fe the
  uncertainty is $3\sigma$.  Bold print indicates our
  \emph{recommended} values used in the fitting of explosive
  nucleosynthesis .
\label{tab:abu}}
\end{table}

Fitting the current SM0313$-$6708 abundances (Table~1) against models
from \citet{HW10} using their {\sc Starfit} algorithm\footnote{see
  also \url{http://starfit.org}; \citep{Chan15}} finds good fits for
progenitor masses in the range $40\,\Msun - 60\,\Msun$ and energies of
$1.5\,\B-1.8\,\B$ with low amounts of mixing, $\lesssim2\,\%$) mixing.
This low amount of mixing is consistent with \citet{HW10} and
hydrodynamic models of Pop III SNe by \citet{Jogg09}.  The best
matching mass is sensitive to the upper limits of the Na and Al
abundances.  Generally, lower values on the upper limits to the
abundances would push the fit toward higher initial masses, stars that
produce a stronger odd-even effect.
The low limits on N in SM0313$-$6708 constrain the mass on the upper
end.

\begin{figure*}
  \centering
\includegraphics[width=0.65\columnwidth]{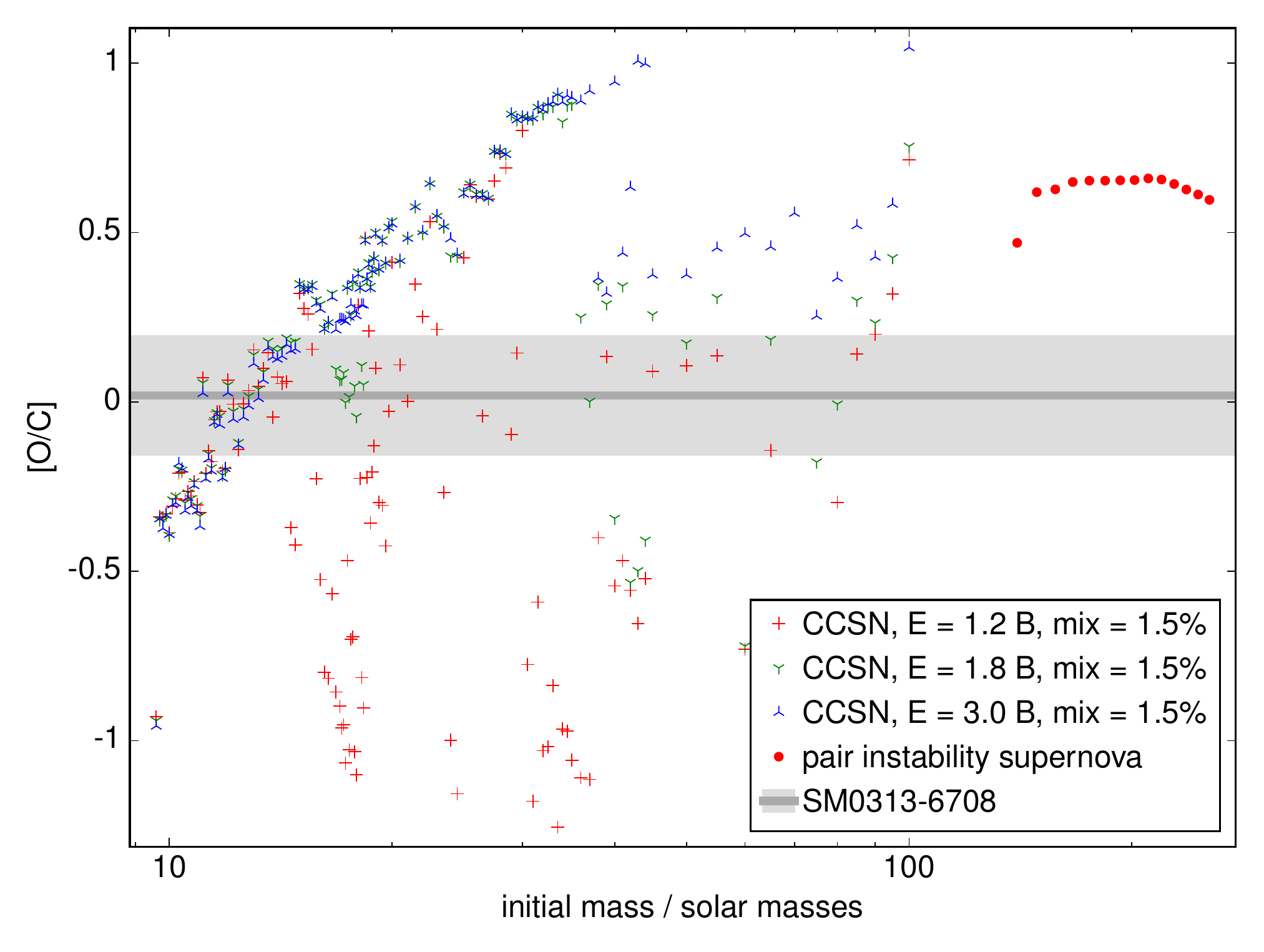}
\includegraphics[width=0.65\columnwidth]{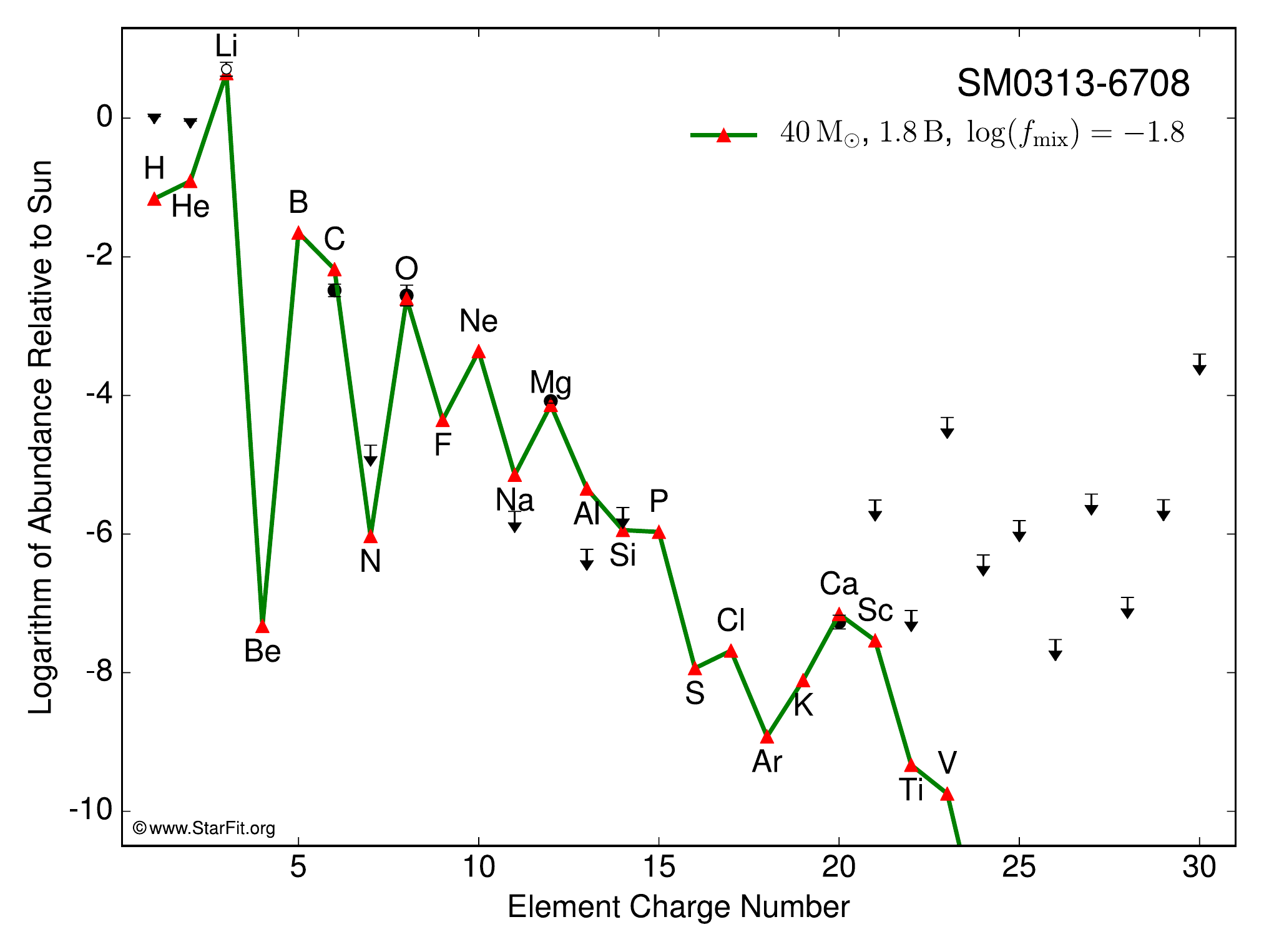}
   \caption{ \textbf{Top:} Range of \SR{O}{C} abundances produced in a
     range of core collapse supernovae for three different explosion
     energies, $1.2\,\B$, $1.8\,\B$, and $3\,\B$ and mixing of
     $1.5\,\%$ (of helium core size; see \citealt{HW10}).  The dark
     gray line and the light gray shading show the derived
     SM0313$-$6708 and its $1\sigma$ error bar of $0.175\,\dex$.
     \textsl{Red dots} show the \SR{O}{C} ratios for Pop III
     pair-instability supernovae \citep{HW02}.  \textbf{Bottom:} The
     yields from stars in the $40\,\Msun - 60\,\Msun$ range provide
     the best fit to the observed abundance pattern of all the
     elements (see the main text).  The $40\,\Msun$ fit is shown here.}
  \label{fig:models}
\end{figure*}

The range of \SR{O}{C} abundances produced in a range of core
collapse supernovae is shown in Fig.~\ref{fig:models} together with a
$40\,\Msun$ fit from {\sc Starfit}.  Significantly, the observed
\SR{O}{C} ratio is lower than what Pop III pair-instability SNe of
\citet{HW02} can produce or what we can produce for high explosion
energies ($E\gtrsim3\,\B$) in more massive stars
($M\gtrsim20\,\Msun$).
The structure and pre-SN evolution of stars with mass above
$15\,\Msun$ is very sensitive to initial mass \citep[e.g.,][]{SW14}
and these stars may experience chaotic nuclear burning.  As a result,
the fraction of the C and O layers that are ejected varies strongly
for a given explosion energy and is rather sensitive to mass and
explosion energy. Currently there is no good first principle
calculations providing reliable predictions for explosion energies for
a given stellar structure for masses above $15\,\Msun$.

Our new 3D LTE values provide definite constraints on the \SR{O}{C}
ratio instead of only an upper limit.  The new lower value of
\SH{Fe} derived here strengthens the constraint \emph{not} to eject
iron group elements from the inner core of the progenitor star.  This
limits the explosion energy and the magnitude of mixing due to
Raleigh-Taylor instabilities during the SN explosion.

\cite{Ishi14} have proposed an alternative scenario to explain the
unique abundances in SM0313$-$6708.  They investigated supernova
yields of metal-free (Population III) stars and find that the high
\SR{C}{Ca} and \SR{C}{Mg} ratios and upper limits of other
elemental abundances are well reproduced with the yields of
core-collapse supernovae (which have normal kinetic energies of
explosion of $1\,\B$) and hypernovae ($10\,\B$) of Population III
$25\,\Msun$ or $40\,\Msun$ stars.  Their best-fit models assume that
the explosions undergo extensive matter mixing and fallback, leaving
behind a black hole remnant.  In these models, Ca is produced by
static/explosive O burning and incomplete Si burning in the Population
III supernova/hypernova, in contrast to our suggestion that the Ca
originates from the hot-CNO cycle during pre-supernova evolution,
which is, remarkably, consistent with the present data.

Although \cite{Ishi14} well fits the observed abundances, their
modelling uses several independent free parameters.  For example,
their models require significantly more mixing than is consistent with
hydrodynamical simulations of Population III stars \citep{Jogg09} and
the fallback is adjusted independently of the explosion energy.
Jetted or asymmetric explosions may be a possibility, however, this
is another free parameter.  In the \citep{HW10} models the
mixing is consistent with hydrodynamical simulations of \citet{Jogg09}
and the fallback is computed self-consistently with the explosion
energy. Fuller discussion of the abundances of SM0313-6708 and
SN modelling will be presented in a future paper.

\section{Conclusions}
The ultraviolet spectrum of SM0313$-$6708 is
dominated by the lines of the OH A-X band which are noticeably
stronger than the CH C-X lines.  The new line lists of Masseron fit
the OH and CH lines very well in both wavelength and strength.  We
could detect no lines of $^{13}$CH, which places a limit of more than
$40$ on $\R{^{12}C}{^{13}C}$, and no lines of NH putting nitrogen at
least two orders of magnitude less than C and O.  This supports there
having been no mixing from the H-burning shell to the surface within
SM0313$-$6708 and there being little nitrogen relative to C and O in
the material from which SM0313$-$6708 formed.  The clear lack of any
metal lines apart from Mg\,I in the UVES ultraviolet spectrum also
enabled us to lower the abundance limits for those elements whose
strong lines are normally seen in this region, including Fe\,I, for
which we now derive an $\SH{Fe}$ upper limit of $-7.52\,\dex$
($3\,\sigma$) from $\langle3D\rangle$ LTE analysis.  We derived an
[$\R{O}C$] ratio of $+0.02$ with a $1\,\sigma$ uncertainty
of $0.175\,\dex$ for SM0313$-$6708, updated abundances measurements
for Mg and Ca, and derived new upper limits for nitrogen, the Fe-group
elements and some of the alpha-elements.  These abundances remain
consistent with a single massive Population III progenitor in the mass
range $40\,\Msun - 60\,\Msun$, low mixing, and modest explosion
energies; the Ca production remains consistent with an origin from
hydrostatic hydrogen burning.  Improved upper limits on
intermediate-mass elements, in particular Na and Al, will further help
constrain the models in the future.  The strong Fe\,II lines in the
vacuum UV between $2340\,\A$ and $2625\,\A$ should allow one to
constrain -- or detect -- the iron abundance of SM0313$-$6708 down to
as low as $\SH{Fe}=-9$.

\section*{Acknowledgments}

AH was supported by an Australian Research Council (ARC) Future
Fellowship (FT120100363).  RC is supported by an ARC Discovery Early
Career Researcher Award (DE120102940).  AC and TM have been supported
by the European Union FP7 programme through ERC grant number 320360.
AF acknowledges support from NSF Career grant AST-1255160.  This
research been supported, in part, by ARC Discovery Project grants
DP120101237 and DP150103294 and NSF grant PHY-1430152 (JINA-CEE).
%

{\it Facilities:} \facility{ESO(UVES)}

\label{lastpage}
\end{document}